\documentclass[aps,pre,twocolumn,superscriptaddress,reprint]{revtex4-1}
\usepackage{amsmath}
\usepackage{amsfonts}
\usepackage{amssymb}
\usepackage{bm}
\usepackage[dvipdfmx]{graphicx}
\usepackage[dvipdfmx]{color}

\begin{document}

\title{A simple relation between frustration and transition points in diluted spin glasses}

\author{Ryoji Miyazaki}
\altaffiliation[Present address: ]{System Platform Research Laboratories, NEC Corporation, Tsukuba, Ibaraki 305-8501, Japan}
\affiliation{Graduate School of Information Sciences, Tohoku University, Sendai 980-8579, Japan}
\author{Yuta Kudo}
\affiliation{Graduate School of Information Sciences, Tohoku University, Sendai 980-8579, Japan}
\author{Masayuki Ohzeki}
\affiliation{Graduate School of Information Sciences, Tohoku University, Sendai 980-8579, Japan}
\affiliation{Institute of Innovative Research, Tokyo Institute of Technology, Kanagawa, 226-8503, Japan}
\author{Kazuyuki Tanaka}
\affiliation{Graduate School of Information Sciences, Tohoku University, Sendai 980-8579, Japan}

\date{\today}

\begin{abstract}
We investigate a possible relation between frustration and phase-transition points in spin glasses.
The relation is represented as a condition of the number of frustrated plaquettes in the lattice at phase-transition points at zero temperature
and was reported to provide very close points to the phase-transition points for several lattices.
Although there has been no proof of the relation, 
the good correspondence in several lattices suggests the validity of the relation 
and some important role of frustration in the phase transitions.
To examine the relation further, 
we present a natural extension of the relation to diluted lattices and verify its effectiveness for bond-diluted square lattices.
We then confirm that the resulting points are in good agreement with the phase-transition points in a wide range of dilution rate.
Our result supports the suggestion from the previous work for non-diluted lattices on the importance of frustration to the phase transition of spin glasses.
\end{abstract}


\maketitle

\section{Introduction}

Spin glasses have been one of the most attractive subjects in statistical mechanics 
and have been extensively investigated for decades~\cite{M.Mezard1987, K.Fischer1993}.
Study of spin glasses, in particular, in infinite dimensions has developed several elaborate concepts and techniques.
For instance, replica symmetry breaking has had a great influence on subsequent studies, 
e.g., structural glasses~\cite{G.Parisi2010, P.Charbonneau2014Jan} and information theory~\cite{H.Nishimori2001, M.Mezard2009}, 
for revealing their complicated energy landscapes.
This success motivates us to tackle a next task that is to establish theories of more realistic models, namely, finite-dimensional spin glasses.
It, however, is a very difficult task.
This is partially because
techniques exploited for the infinite-dimensional models 
are not so useful in finite dimensions, 
while it is still difficult to obtain conclusive proofs with numerical simulations.

The gauge transformation~\cite{H.Nishimori2001} has been utilized as a tool for analytically investigating finite-dimensional spin glasses~\cite{S.Morita2006, H.Nishimori2007, M.Ohzeki2008, M.Ohzeki2009Feb, M.Ohzeki2009JulJPA, M.Ohzeki2012Jun, M.Ohzeki2012Nov, M.Ohzeki2013, M.Ohzeki2015Feb}.
A consequence of this approach is the conjecture on the verticality of a phase boundary~\cite{H.Nishimori1986}.
The conjecture states that 
the phase boundary between the ferromagnetic and another phases at low temperatures does not depend on temperature 
but is determined only by geometrical properties.
In other words, 
the phase boundary is a vertical line in the $p$--$T$ plane, 
where $p$ and $T$ denote the ratio of antiferromagnetic bonds of spins and temperature, respectively.
This conjecture was denied by subsequent detailed studies~\cite{A.Honecker2001, S.deQueiroz2006, M.Ohzeki2009Feb, M.Achilles2000, C.Wang2003, C.Amoruso2004, K.Fujii2012, N.Jinuntuya2012}.
The established phase boundary, however, is almost vertical.
This fact implies that geometrical properties take a primary role for the phase transition, 
even though the conjectured relation does not exactly hold.
Note that 
the importance of revealing the property of this phase transition is not limited in the study of spin glasses.
It can influence on the study of quantum computation.
Indeed,
this phase boundary can be interpreted to give the error-correction threshold 
for topological quantum error-correction codes~\cite{A.Kitaev2003, E.Dennis2002}.

Another possible relation 
between a geometrical property and this phase transition was reported by one of the authors without utilizing the gauge transformation~\cite{R.Miyazaki2013Jul}.
The relation is represented as a condition on a quantity concerning frustration~\cite{G.Toulouse1977, S.Kirkpatrick1977, J.Vannimenus1977} in the lattice.
The condition gives a very close point to the phase-transition point at zero temperature.
The good correspondence is found in several two-dimensional lattices and hierarchical lattices~\cite{N.Berker1979}.
Moreover,
the condition for the Sherrington-Kirkpatrick model~\cite{D.Sherrington1975} exactly gives the replica symmetry solution for its transition point at zero temperature.
Unfortunately, 
we have no proof that 
this agreement is not just an accidental one.
The above instances, however, allow us to expect some important role of frustration in the phase transition.

Recently, some of the authors extended this argument to bond-diluted lattices~\cite{M.Ohzeki2018}.
Their method mainly follows the above one for the non-diluted lattices.
Resulting points from their method qualitatively agrees with the correct phase transition points.
However, the method was not exactly executed.
They instead used an additional ansatz to complete the calculations  
because of the difficulty due to the inhomogeneity in the diluted lattices.
A natural question is 
whether the good correspondence is also found by the canonical extension without the ansatz 
or is just caused by the ansatz.

In this paper, 
we examine the natural extension of the method for non-diluted lattices~\cite{R.Miyazaki2013Jul} to bond-diluted lattices without any extra ansatz.
The next section gives the prescription of the method.
We describe two natural ways of extension of the method for non-diluted lattices to the diluted case.
The effectiveness of the method is verified for the diluted square lattice in Sec.~\ref{sec:square}.
We apply the method with perturbative analysis expanded from the non-diluted case and numerical calculations.
The obtained points are compared with the correct phase-transition points.
We summarize and discuss our results in Sec.~\ref{sec:summary}.

\section{Prescription}
\label{sec:prescription}

We investigate $\pm J$ Ising spin glass~\cite{H.Nishimori2001}, 
defined by
\begin{equation}
H = - \sum_{\langle i, j \rangle} J_{ij}\sigma_i \sigma_j, 
\end{equation}
where $\langle i, j \rangle$ denotes a pair of nearest-neighbor sites on a lattice.
The coupling constants $J_{ij}$ for spin $i$ and $j$ are taken from an independent, identical distribution 
$P(J_{ij}) = p \delta(J_{ij} + J) + (1-p)\delta (J_{ij}-J)$ with $J>0$, 
where $\delta(\cdot)$ is the Dirac delta function.
Note that we define $p$ as the probability that a bond of spins is an antiferromagnetic one.
Ising spin $\sigma_i$ takes $1$ or $-1$.
This model has been extensively used as an elementary model of spin glasses in finite dimensions~\cite{H.Nishimori2001}.
We mainly consider the model on two-dimensional lattices.

We focus on frustration for plaquettes~\cite{G.Toulouse1977, S.Kirkpatrick1977, J.Vannimenus1977}.
A plaquette is an elementary loop of edges on a lattice, 
which cannot be divided into multiple sub-loops.
For example, a plaquette on a square lattice is a square composed of four edges.
When a plaquette has an odd number of antiferromagnetic couplings, 
there is no spin configuration that all bonds in the plaquette take the lower energy state.
Consequently, there is frustration at the plaquette.
Such a plaquette is called frustrated plaquette.
The average number of frustrated plaquettes on a lattice over the bond distribution plays a central role in the argument below.
That average is calculated as
\begin{equation}
N_\text{fra}(p) = \left\langle \sum_{c}\frac{1}{2} \left( 1-\prod_{\langle i, j \rangle \in c} \frac{J_{ij}}{J} \right) \right\rangle^\text{af}_p, 
\end{equation}
where $c$ are indices for plaquettes, 
and $\langle \cdot \rangle^\text{af}_p$ denotes the average over the antiferromagnetic-bond distribution for $p$.
The coupling constants are independent of each other, 
and hence we can rewrite the function as
\begin{align}
N_\text{fra} (p) &= \sum_n N_\text{pla}^{(n)} f_n(p), \label{eq:N_fra(p)_2} \\
f_n(p) &= \frac{1}{2} \left[1 - \left( 1-2p \right)^n \right], \label{eq:f_n}
\end{align}
where $N_\text{pla}^{(n)}$ is the number of plaquettes composed of $n$ edges on the lattice.
The function $f_n(p)$ gives the probability that 
a plaquette composed of $n$ edges is frustrated.
For the square lattice, the sum reduces to the single term for $n=4$.
This expression also concerns lattices with multiple types of plaquettes, e.g., the Kagom\'e lattice.

One of the authors focused on the function~\cite{R.Miyazaki2013Jul} defined by
\begin{equation}
v(p) = \frac{d N_\text{fra}(p)}{dp} \left( \frac{d N_\text{af}(p)}{dp} \right)^{-1}. \label{eq:v(p)}
\end{equation}
Here $N_\text{af}(p)$ is the average number of antiferromagnetic bonds over the bond distribution for $p$, 
calculated as $N_\text{af}(p) = p N_\text{edg}$, 
where $N_\text{edg}$ is the number of edges in the lattice.
He reported~\cite{R.Miyazaki2013Jul} that 
the condition $v(p) = 1$ gives a value of $p$ that is close to the phase transition point for the model at zero temperature.
For instance, 
the value $p \simeq 0.1031$ yielded from the condition for the square lattice is very close to the actual phase-transition point 
numerically obtained as $p = 0.1033(1)$~\cite{K.Fujii2012} or $0.1045(11)$~\cite{N.Jinuntuya2012}.
The good correspondence is found in several two-dimensional lattices and hierarchical lattices~\cite{N.Berker1979}.
Interestingly, 
the condition for the Sherrington-Kirkpatrick model~\cite{D.Sherrington1975} exactly gives the replica symmetry solution for its phase-transition point at zero temperature.

We extend the above argument to apply to $\pm J$ Ising spin glass on bond-diluted lattices.
Each edge in the lattices is absent with probability $q$.
The probability that a bond is an antiferromagnetic one thus turns to $(1-q)p$.
We introduce two ways of extension to this case.
The first one generalizes the function in Eq.~(\ref{eq:N_fra(p)_2}) 
as the average number of frustrated plaquettes over diluted lattices as well as the antiferromagnetic-bond distributions, namely 
\begin{equation}
N_\text{fra}(q, p) = \sum_n \left\langle N_\text{pla}^{(n)} \right\rangle^\text{di}_q f_n(p), \label{eq:N_fra(q,p)}
\end{equation}
where $\langle \cdot \rangle^\text{di}_q$ denotes the average over the diluted-bond distribution for $q$.
We have utilized the fact that 
$f_n(p)$ is not affected by the bond dilution.
Accordingly, we define a generalized function of $v(p)$ in Eq.~(\ref{eq:v(p)}) by
\begin{equation}
v(q,p) = \frac{\partial N_\text{fra}(q,p)}{\partial p} \left( \frac{\partial N_\text{af}(q,p)}{\partial p} \right)^{-1}. \label{eq:v(q,p)}
\end{equation}
Here $N_\text{af}(q,p)$ is the average number of antiferromagnetic bonds over the bond distribution for $q$ and $p$, 
calculated as $N_\text{af}(q,p) = (1-q)p N_\text{edg}$, 
where $N_\text{edg}$ is the number of edges in the lattice without dilution.
The condition $v(p) = 1$ is generalized as $v(q,p) = 1$ for the bond-diluted lattices.
For the other extension 
we calculate the average number of frustrate plaquettes and antiferromagnetic bonds 
over only the antiferromagnetic-bond distribution on a given bond-diluted lattice.
Equation~(\ref{eq:v(p)}) then gives the function $v(p)$ for the lattice.
We consider a bond-diluted lattice for this extension, 
whereas we took the average over bond-diluted lattices for $q$ for the first extension. 
The second extension could give different solutions of $v(p) = 1$ for different diluted lattices.
However, 
if the variance of obtained solutions is small for the lattices for $q$, 
we would find a typical value of $p$ for the condition $v(p) = 1$ for the diluted lattices.
The possible typical one is regarded as the solution obtained from our method for $q$.
In addition, 
we can consider a minor change of this extension in estimation of the typical solution;
we estimate the average of the function $v(p)$ over bond-diluted lattices 
and obtain the solution of $v(p) = 1$ for the averaged function instead of the average of solutions themselves over different lattices.
Hereafter, we examine whether the two procedures of extension give $p$ close to the correct phase transition point of the model on diluted lattices for $q$ observed with varying $p$ at zero temperature.

\section{Square lattice}
\label{sec:square}

\subsection{Perturbative calculations}

We first restrict our interest to the systems in which the number of lacked edges is small 
and obtain its expansion in terms of $q$.
Motivated by the fact that 
$p$ for the condition $v(p) = 1$ for the square lattice is extremely close to the phase-transition point~\cite{R.Miyazaki2013Jul}, 
we analyze the model on the square lattice.
Here, we only attempt the first way of extension, 
where we calculate $v(q,p)$ in Eq.~(\ref{eq:v(q,p)}), 
because it is intractable to analytically obtain the solutions with the second extension.
The second one will be examined with numerical calculations in Sec.~\ref{sec:numerical}.
It should be noted that 
$N_\text{edg}$ and $N_\text{pla}$ used below denote the numbers of edges and plaquettes, respectively, for the lattice without dilution.

\begin{figure}
	\begin{center}
	\includegraphics[width=\columnwidth]{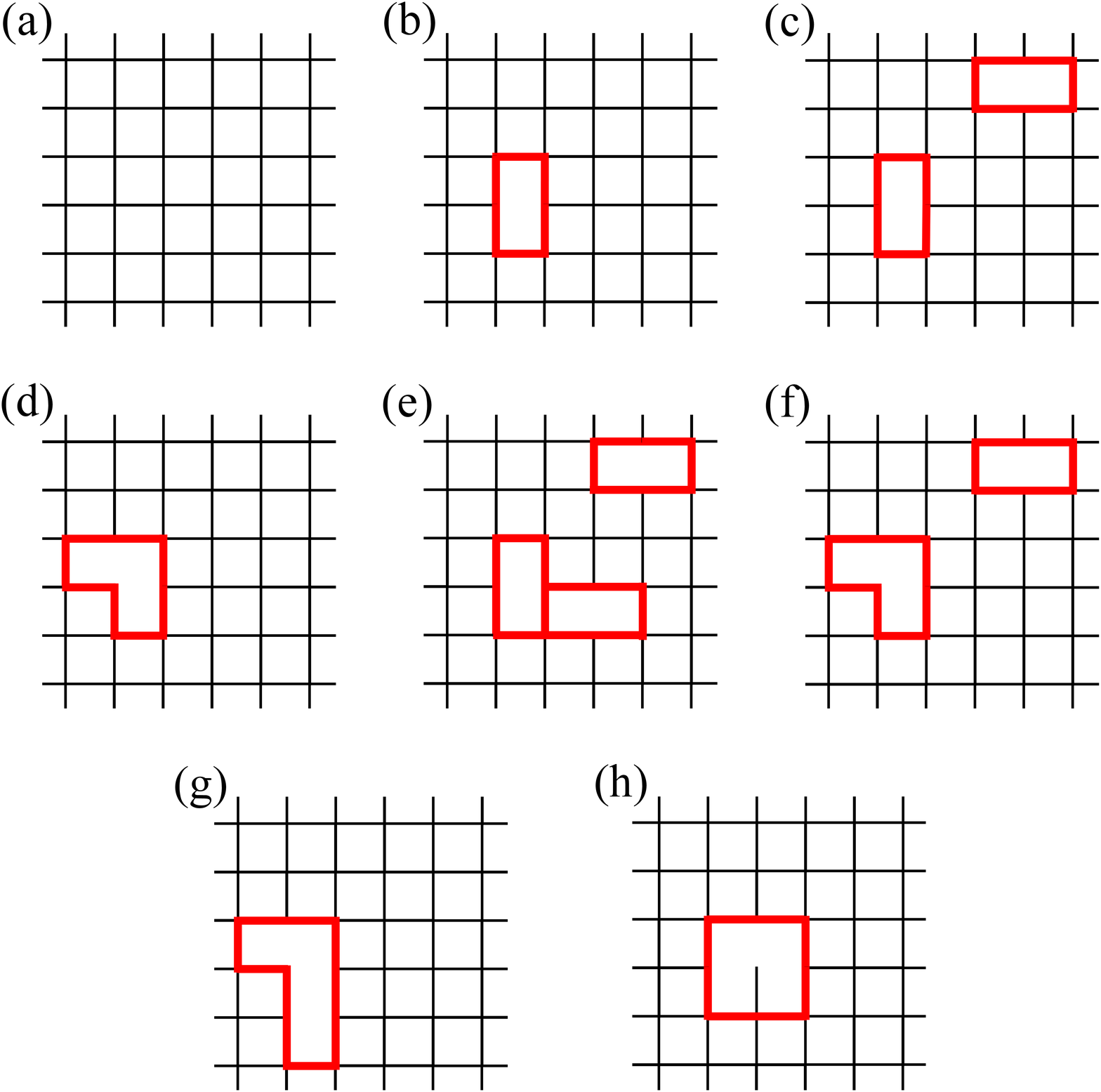}
	\end{center}
	\caption{
		Examples of square lattices without/with dilutions treated in calculations of $N_\text{fra}(q,p)$ for small $q$.
		Plaquettes, which are not composed of four edges, generated by removing edges are highlighted with thick red lines.
		(a) The square lattice without dilution.
		Examples of square lattices lacking (b) an edge, (c), (d) two edges, and (e)--(h) three edges.
		The examples are distinguished by the number of lacked edges and the number of edges for generated plaquettes.
		}
	\label{fig:square_lattices}
\end{figure}

\begin{figure}
	\begin{center}
	\includegraphics[width=\columnwidth]{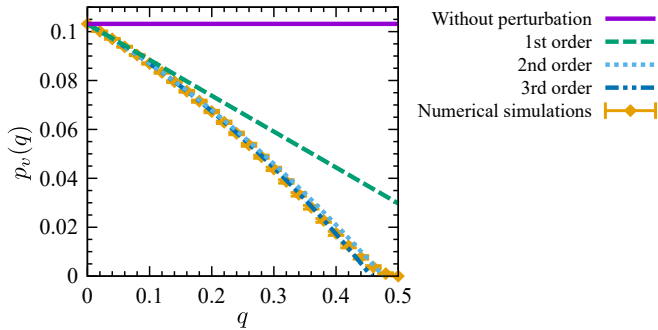}
	\end{center}
	\caption{Solutions $p_v(q)$ obtained with the perturbative calculations of the first extension and with the numerical calculations of the second one.
		The $n$th-order perturbative solution is a function of $q$ 
		taking into account lower order terms than $q^{n+1}$ given in Eq.~(\ref{eq:p_v(q)_perturb}) for $n=0$ (without perturbation), 1, 2, and 3.
		The numerical simulations are done for the lattices generated by removing edges from the square lattice of $L \times L$ units, 
		where $L$ for the result shown here is 128.
		The numerical solutions are estimated by averaging $10^4$ instances.
		}
	\label{fig:p_q_analytic}
\end{figure}

We obtain $N_\text{fra}(q, p)$ in Eq.~(\ref{eq:N_fra(q,p)}) for terms up to $q^n$ 
by considering lattices in which the number of lacked edges is smaller than $n+1$.
This is because the probability that a lattice lacks $n$ edges is $q^n(1-q)^{N_\text{edg}-n}$, 
and because $\langle N_\text{pla}^{(n)} \rangle^\text{di}_q$ in $N_\text{fra}(q, p)$ is the average number of $n$-edge plaquettes over those lattices.
Figure \ref{fig:square_lattices} shows examples of square lattices removed 1, 2, or 3 edges.
Note that there can exist edges which do not belong to any plaquette 
and thus	 do not contribute to frustration.
For instance, 
the edge in the 8-edge square on the lattice shown in Fig.~\ref{fig:square_lattices}~(h) 
does not belong to any loop of edges.
As an example of computing $N_\text{fra}(q,p)$
let us consider a lattice lacking an edge as shown in Fig.~\ref{fig:square_lattices}~(b).
The probability that such a lattice is realized is $q(1-q)^{N_\text{edg}-1}$.
The number of positions at which an edge is absent is $N_\text{edg}$.
By removing an edge from the primary square lattice, 
the number of four-edge plaquettes reduces to $N_\text{pla}-2$, 
while a six-edge plaquette is generated.
The contribution of such lattices to $N_\text{fra}(q,p)$ is thus 
$q (1-q)^{N_\text{edg}-1} N_\text{edg} [ \left( N_\text{pla} - 2 \right) f_4(p) + f_6(p) ]$.
Taking into account the lattices removed 1, 2, or 3 edges, 
we obtain
\begin{equation}
\begin{split}
N_\text{fra} & (q, p) \\
=& (1-q)^{N_\text{edg}} N_\text{pla} f_4(p) \\
&+ q (1-q)^{N_\text{edg}-1} N_\text{edg} [ \left( N_\text{pla} - 2 \right) f_4(p) + f_6(p) ] \\
&+ q^2 (1-q)^{N_\text{edg}-2} N_\text{edg} \\
&\times \bigg\{ \frac{N_\text{edg} - 7}{2} [ ( N_\text{pla} - 4 ) f_4(p) + 2f_6(p) ] \\
& \hspace{20pt} + 3[ ( N_\text{pla} - 3 ) f_4 (p) + f_8(p) ] \bigg\} \\
&+ q^3 (1-q)^{N_\text{edg}-3} N_\text{edg} \\
& \times \bigg\{ [ 2(N_\text{edg} - 12 ) + 14 (N_\text{edg} - 13 ) \\
& \hspace{20pt} + (N_\text{edg} - 23 )(N_\text{edg} - 14 ) ] \\
& \hspace{20pt} \times \frac{1}{3!} [(N_\text{pla}-6)f_4(p) + 3 f_6(p) ] \\
& \hspace{20pt} + 3 (N_\text{edg} - 10) [(N_\text{pla}-5) f_4(p) + f_6(p) + f_8(p)] \\
& \hspace{20pt} + 9 [(N_\text{pla}-4) f_4(p) + f_{10}(p)] \\
& \hspace{20pt} + 2 [(N_\text{pla}-4) f_4(p) + f_8(p)]\bigg\} + O(q^4) \\
=& N_\text{pla} f_4(p) + N_\text{edg} [-2 f_4(p) + f_6(p)]q \\
&+ 3N_\text{edg} [f_4(p) - 2f_6(p) + f_8(p)]q^2 \\
&+ N_\text{edg} [-2f_4(p) + 15f_6(p) -22f_8(p) + 9f_{10}(p)]q^3 \\
& + O(q^4).
\end{split}
\end{equation}
Substituting this into Eq.~(\ref{eq:v(q,p)}), we then have
\begin{equation}
\begin{split}
v(q,p)
=& 2r^3 \big[ 1 - 3\left(1-r^2\right)q + 3 \left( 1-5r^2+4r^4 \right)q^2 \\
& \hspace{20pt} + \left( -1 + 30r^2 -76 r^4 + 45r^6\right)q^3 \big] + O(q^4)
\end{split}
\end{equation}
where $r = 1-2p$.
We have used a relation $N_\text{pla} = N_\text{edg}/2$ for the square lattice.
The solution $p_v(q)$ of $v(q,p) = 1$ for $q$ is expanded in terms of $q$ as
\begin{equation}
\begin{split}
p_v(q)
=& \frac{1}{2} - 2^{-4/3} - \left[ 2^{-4/3} -2^{-2} \right] q  - \left[ 2^{-4/3} -\frac{1}{4} \right] q^2 \\
&  - \left[ \frac{11}{6} \times 2^{-4/3} - \frac{4}{3}\times2^{-5/3} - \frac{1}{4} \right] q^3 + O(q^4).
\end{split} \label{eq:p_v(q)_perturb}
\end{equation}
The solutions containing the terms up to $q^n$ for $n = 0$, 1, 2, and 3 are drawn in Fig.~\ref{fig:p_q_analytic}, 
where a result of the second extension given in Sec.~\ref{sec:numerical} is also shown for comparison.

\subsection{Numerical calculations}
\label{sec:numerical}

We run numerical simulations of the second extension, 
where the solution of $v(p) = 1$ for each bond-diluted lattice is estimated.
We first generate a bond-diluted lattice under the periodic boundary condition in which
an edge is lacked with probability $q$ 
and then count plaquettes and edges.
We do not consider whether bonds in the lattice are ferromagnetic ones or antiferromagnetic ones, 
since this matter concerns only the $p$ dependence of $N_\text{fra}(p)$ 
that is already determined by $f_n(p)$ given in Eq.~(\ref{eq:f_n}).
We then obtain the function $v(p)$ and the solution of $v(p) = 1$ for the lattice.
Sampling solutions for a number of diluted lattices for $q$ by this way, 
we estimate the average and variance of the solutions.
The average is also denoted by $p_v(q)$ for simplicity.

Figure~\ref{fig:p_q_analytic} shows the plot of the estimated solutions as a function of $q$. 
The perturbative solutions based on the first extension are also displayed for comparison.
The square lattice before the dilution has $L \times L$ units (squares), 
where $L$ for the result shown in Fig.~\ref{fig:p_q_analytic} is $128$. 
Solutions of $v(p) = 1$ are sampled from $10^4$ lattices generated from the distribution for $q$.
The variance of the solutions over different lattices is small.
The averaged value is thus regarded as the probable solution for $q$ obtained with the second extension of our method.
In addition, 
the obtained numerical solution of the second extension is in good agreement with the perturbative solutions of the first extension for small $q$.
In particular, the numerical solution and the third-order perturbative solution show good correspondence for $q \le 0.4$.
This result demonstrates that
both the ways of extension lead to almost identical solutions.
The perturbative solutions, however, do not exhibit the non-monotonic behavior found in the numerical ones at $q > 0.4$, 
where the perturbative analysis expanded from $q=0$ would be unreliable.
We should remark that
the curve of the numerical solutions converges to 0 with $q$ approaching 0.5, 
which agrees with the exact phase-transition point at $q=0.5$.

\begin{figure}
	\begin{center}
	\includegraphics[width=\columnwidth]{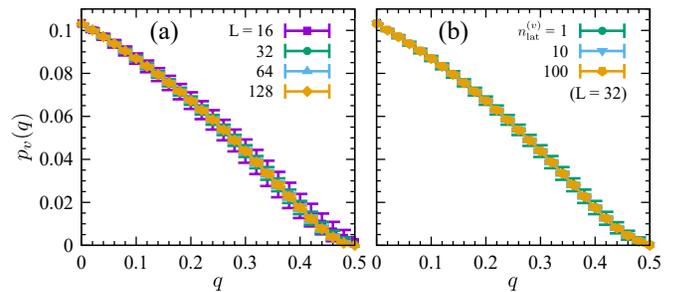}
	\end{center}
	\caption{Solutions $p_v(q)$ of $v(p) = 1$ estimated from $10^4$ instances of numerical simulations based on the second extension.
	(a) The solutions for $L = 16$, 32, 64, and 128.
	(b) The solutions for the averaged function over $n_\text{lat}^{(v)}$ lattices for $n_\text{lat}^{(v)} = 1$, 10, and 100 for $L = 32$.
		}
	\label{fig:p_q_numerical}
\end{figure}

The size $L$ dependence of the solutions is shown in Fig.~\ref{fig:p_q_numerical}~(a).
We find no definite difference in the average of the solutions between investigated $L$ 
except for $q \ge 0.4$, 
where a slight decrease is observed with increasing $L$. 
We therefore expect that the finite-size effect of our solutions is small.
On the other hand, 
the variance of the solutions clearly decreases as $L$ increases.
As mentioned above, 
the average of the solutions over bond-diluted lattices agrees well with the solutions of the first extension, 
where we obtained the solutions with the average number of frustrated plaquettes.
This finding and the decrease of the variance with increasing $L$ suggest that
the small variance of the numerical solution originates from the typicality of the number of frustrated plaquettes
that could be involved in the self-averaging property~\cite{H.Nishimori2001} of the system.

Using a rather small lattice ($L=32$), 
we also execute the other procedure of estimation of the typical solution mentioned in the end of Sec.~\ref{sec:prescription}, 
where we compute the solutions of the averaged $v(p)$ over $n_\text{lat}^{(v)}$ lattices. 
To observe the variance of the resulting solutions, they are sampled $10^4$ times.
This estimation for $n_\text{lat}^{(v)} = 1$, hence, corresponds to the above method 
the result of which is shown in Fig.~\ref{fig:p_q_numerical}~(a) ($L = 32$).
Figure~\ref{fig:p_q_numerical}~(b) displays the average of the obtained solutions with error bars over $10^4$ samples 
for $n_\text{lat}^{(v)} = 1, 10, 100$.
Increasing $n_\text{lat}^{(v)}$ does not make any definite differences in the average of the solutions 
but just suppresses the fluctuation of the solutions. 
Therefore, we use the result for $n_\text{lat}^{(v)} = 1$ as the solution of our method for the bond-diluted lattices.

\subsection{Comparison with the minimum-weight perfect-matching algorithm}

We compare the obtained solution of $v(p) = 1$ with the correct phase-transition point.
The latter has been already estimated 
in the context of the quantum error correction for the surface code with loss
by using the minimum-weight perfect-matching (MWPM) algorithm~\cite{T.Stace2009, *T.Stace2010}.
We, however, performed the similar calculations in a number of points of $q$,  
because we need detailed illustration of the $q$ dependence of the critical point. 
We followed the treatment of the diluted lattices 
as well as the system size, $L =16$, 24, 32, 
and the number of instances of diluted lattices, $5\times 10^4$, in Ref.~\cite{T.Stace2009, *T.Stace2010} 
and the finite-size scaling ansatz in Ref.~\cite{C.Wang2003}.

\begin{figure}
	\begin{center}
	\includegraphics[width=0.9\columnwidth]{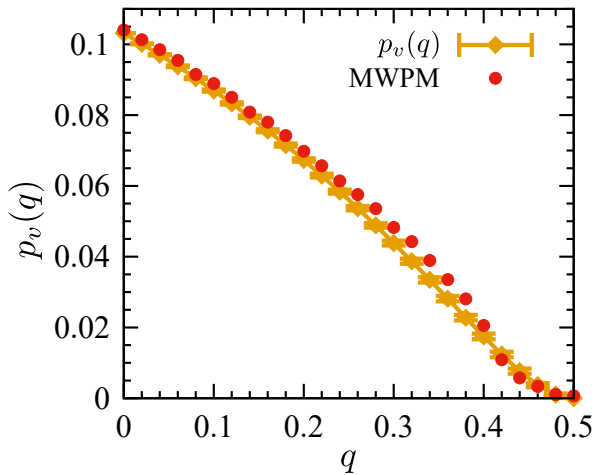}
	\end{center}
	\caption{Solutions $p_v(q)$ of $v(p) = 1$ for $L = 128$ (the same one as shown in Fig.~\ref{fig:p_q_analytic}) 
				and the phase-transition points estimated with MWPM~\cite{T.Stace2009, *T.Stace2010}.
		}
	\label{fig:p_q_mwpm}
\end{figure}

The average $p_v(q)$ of solutions of $v(p) = 1$ is in good agreement with the obtained critical values $p_c$ 
in the whole range of $q$ ($0 \le q \le 0.5$) except for $q \simeq 0.34$, 
as shown in Fig.~\ref{fig:p_q_mwpm}.
The correspondence at $q=0$ previously found~\cite{R.Miyazaki2013Jul} is reproduced.
Interestingly, 
the slope of $p_v(q)$ at $q=0$ is also very similar to that of $p_c$.
Moreover, they remain almost identical curves for $q \le 0.14$. 
This finding demonstrates that
our method effectively captures the $q$ dependence of the true phase-transition point for small $q$ at least.
For larger $q$, 
$p_v(q)$ departures from $p_c$ 
and takes a little smaller value.
The difference between them takes its maximum around $q \simeq 0.34$, 
but it is still small.
For $q \ge 0.34$, $p_c$ decreases more rapidly than $p_v(q)$, 
and they take similar values again at $q\ge 0.4$.
This agreement is owed to that 
the non-monotonic behavior for $q \ge 0.4$ in $p_v(q)$, mentioned in Sec.~\ref{sec:numerical}, 
appears also in $p_c$.
Both the curves finally converge to 0 with $q$ approaching 0.5.
This good correspondence in the range of $q$ implies a scenario that 
our simple method could give some approximate location of the phase-transition point even for the diluted lattices, 
although we have not been able to directly derive their relationship.

The non-monotonic behavior in the curve of $p_c$ was already reported in the previous work~\cite{T.Stace2009, *T.Stace2010}.
This was attributed to a finite-size effect~\cite{T.Stace2009, *T.Stace2010} 
because of the fact that 
the largest plaquette occupies approximately half of the primary square lattice 
in the range for the non-monotonic behavior.
More precisely, 
the threshold values of $q$, say $q_\text{scale}(L)$, 
at which the largest plaquette occupies half of the lattice were estimated as a function of $L$.
The non-monotonic behavior indeed appeared for larger $q$ than $q_\text{scale}(L)$ for $L$ 
used in estimating the critical points~\cite{T.Stace2009, *T.Stace2010}.
Hence, the non-monotonic behavior in our curve for $p_c$ would be regarded as a signal of the finite-size effect. 
This argument derives that
for the lattice of $L = 128$, 
which is used for our numerical analysis of $v(p)$ and is much larger than that for the estimation with MWPM ($L = 16$, 24, and 32), 
the finite-size effect caused by the occupation of the large plaquettes is not supposed to be observed for $q\le 0.46$ at least~\cite{T.Stace2009, *T.Stace2010}.
$p_v(q)$, however, exhibits the non-monotonic behavior in $q \le 0.46$.
This fact implies that
the non-monotonic behavior of $p_v(q)$ is not due to the occupation of the large plaquettes.
If our method based on $v(p)$ is effective to approximately predict the phase-transition point even for large $q$, 
our result supports that 
the non-monotonic behavior in the estimated $p_c$ was accidentally identified to the finite-size effect 
but is a nontrivial feature of this phase transition.

\section{Summary and discussion}
\label{sec:summary}

We presented a possible relation 
between frustration and the phase transition of Ising spin glasses on bond-diluted lattices.
The relation is represented as the correspondence of points obtained by a simple method concerning frustration and the phase-transition points at zero temperature observed with varying the ratio of antiferromagnetic bonds.
The method is based on extension of a previous one for lattices without dilution~\cite{R.Miyazaki2013Jul}.
We calculate $v(q,p)$ defined by Eq.~(\ref{eq:v(q,p)}) using averaged quantities over diluted lattices 
or $v(p)$ defined by Eq.~(\ref{eq:v(p)}) for each diluted lattice.
Both the two functions concern the derivative of the number of frustrated plaquettes with respect to the number of antiferromagnetic bonds in the lattice.
This extension is more natural than another one previously proposed with an additional ansatz~\cite{M.Ohzeki2018}.
Motivated by the work for non-diluted lattices~\cite{R.Miyazaki2013Jul}, 
where the condition that the obtained function is equal to unity leads to an approximate location of the phase-transition point, 
we applied the extended method to the diluted square lattice. 
Consequently, 
we found that 
both the two ways of extension typically give almost identical result
and that the obtained curve as a function of $q$ is close to the correct phase boundary in the range $0 \le q \le 0.5$. 
A remarkable feature of the curve is non-monotonic behavior in $q>0.4$.
Although the similar feature found in the correct phase boundary in the same range was attributed to a finite-size effect~\cite{T.Stace2009, *T.Stace2010}, 
our case is not simply regarded as the finite-size effect, 
since we investigated larger lattices 
which are not supposed to exhibit the finite-size effect for given $q$.

Our scheme provides close points to the phase-transition points even for diluted lattices.
We, however, never propose it as a method to obtain the phase transition points, 
since the reason for the good correspondence has not been revealed.
We need further investigation to clarify 
whether this agreement is reasonable or not.
We will examine other diluted lattices as the next task.
If the good agreement is not an accident, 
our result suggests that 
geometrical properties can almost fully determine the phase-transition points.
Moreover, 
the non-monotonic behavior in the phase boundary might be a genuine feature of the phase transition.

\begin{acknowledgments}
This research is partially supported by JSPS KAKENHI Grant No. 18H03303 and 19H01095, and the JST-CREST (No.JPMJCR1402) for Japan Science and Technology Agency.
\end{acknowledgments}

\bibliography{ref_frustration_dilution}

\end{document}